\documentclass{webofc}
\usepackage[varg]{txfonts}   
\begin{document}
\title{Fluctuating shapes of the fireballs in heavy-ion collisions}
%
%

\author{\firstname{Boris} \lastname{Tom\'a\v{s}ik}\inst{1,2}\fnsep\thanks{\email{boris.tomasik@cern.ch}} 
	\and
        \firstname{Jakub} \lastname{Cimerman}\inst{2} 
        \and
        \firstname{Renata} \lastname{Kope\v{c}n\'a}\inst{3}
        \and
        \firstname{Martin} \lastname{Schulc}\inst{4}
}

\institute{Univerzita Mateja Bela, Tajovsk\'eho 40, 97401 Bansk\'a Bystrica, Slovakia
\and
FNSPE, \v{C}esk\'e vysok\'e u\v{c}en\'i technick\'e v Praze, B\v{r}ehov\'a 7, 11519 Praha 1, Czechia
\and
Physikalisches Institut, Rupprecht-Karls Universit\"at Heidelberg, Im Neuenheimer Feld 226,\\ 69120 Heidelberg, Germany
\and
Research Centre \v{R}e\v{z} Ltd, 250 68 Husinec-\v{R}e\v{z} 130, Czechia 
 }

\abstract{%
We argue that energy and momentum deposition from hard partons into quark-gluon plasma induces an important 
contribution to the final state hadron anisotropies. We also advocate a novel method of Event Shape Sorting which 
allow to analyse the azimuthal anisotropies of the fireball dynamics in more detail. A use of the method in femtoscopy
is demonstrated. 
}
\maketitle


\section{Motivation}
\label{intro}

In ultrarelativistic heavy-ion collisions, numbers of produced hadrons are so large that empirical distributions of their 
azimuthal angles can reasonably be constructed.  
Large anisotropies exist in those distributions. Moreover, these anisotropies are different in every collision event. 
The usual paradigm is that these anisotropies result from the response of the hot matter to the inhomogeneities 
within its initial state. This response is determined by features like the Equation of State (EoS) or the transport 
coefficients. Those we would like to extract from the response. The caveat is that the initial conditions are largely 
unknown and not directly accessible through a measurement. 

In this overview talk we bring up two issues which concern this standard interpretation of the fireball evolution. 

In the first part we argue that the final state anisotropies can also be produced by the momentum and energy deposition 
from hard partons traversing the deconfined matter \cite{Schulc:2014jma,Tachibana:2014lja}.
This source of anisotropy, which is not present in the initial conditions, is relevant in collisions 
at the LHC energy. 

Since usually many events are summed up in order to obtain better statistic, some anisotropies may be averaged 
out in such summations. The method of Event Shape Sorting \cite{Kopecna:2015fwa}, introduced in 
the second part of this talk, helps to organise the event averaging in such a way that the final state anisotropies 
survive better. Such events with a richer structure of the azimuthal angle distribution might better 
help to reconstruct the properties of the hot matter from the response to initial inhomogeneities.


\section{Bulk anisotropic flow from hard partons}
\label{s:part1}

In nuclear collisions at LHC energies, many hard partons are produced in the initial interactions 
of the incoming partons from the two nuclei. The share of energy out of all energy deposited in 
a collision is larger than at RHIC or SPS. 
Shortly after they are produced, the deconfined medium fills the space around them. Hence, they 
evolve in this quark-gluon plasma (QGP) and loose energy and momentum in favour of the bulk. 
Depending on the total energy of the parton and the size of the energy loss, this process spans over 
some time. During this time the deposited momentum and energy lead to production of collective 
mechanical effects in the QGP: wakes and Mach cones 
\cite{Satarov:2005mv,CasalderreySolana:2004qm,Neufeld:2008fi}. Most important for us are the streams 
in the wakes which carry the momentum of the original hard parton \cite{Schulc:2013kra}.

The resulting flow anisotropies are different in each event. A naive expectation would be that they 
are averaged out if many events are summed up, because hard partons are produced isotropically in the
transverse plane. It turns out that this is not true. They are correlated with the geometry of a non-central 
collision. Two streams are more likely to meet each other if they are produced in the direction 
perpendicular to the reaction plane. Then they (partially) cancel each others momentum. Thus 
more bulk flow is induced in the direction of the impact parameter. This is a positive contribution 
to the elliptic flow due to pressure gradient anisotropy in the initial state. 

In order to study this effect we implemented source terms in 3D ideal hydrodynamic simulation \cite{Schulc:2014jma}. 
The used Equation of State was parametrised from lattice QCD results \cite{Huovinen:2009yb}. Initial 
conditions were smooth, calculated with optical Glauber model and the rapidity profile of the fluid was 
flat. Hard partons were introduced into this environment at the beginning of its evolution. They are 
produced as back-to-back pairs (due to transverse momentum conservation) with fluctuating number of 
pairs and power-law distribution of the $p_t$'s. 

\begin{figure*}
\centering
\includegraphics[width=0.495\textwidth,clip]{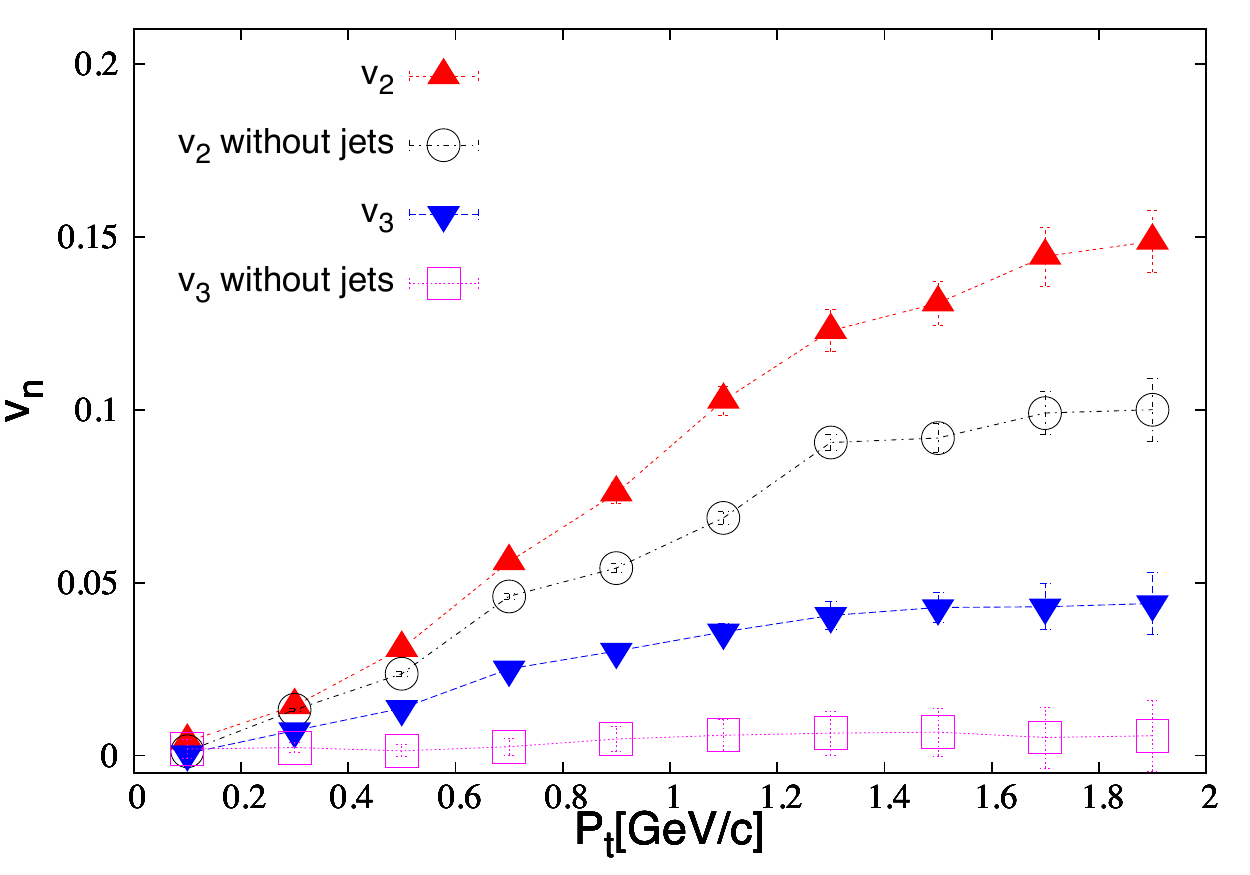}
\includegraphics[width=0.495\textwidth,clip]{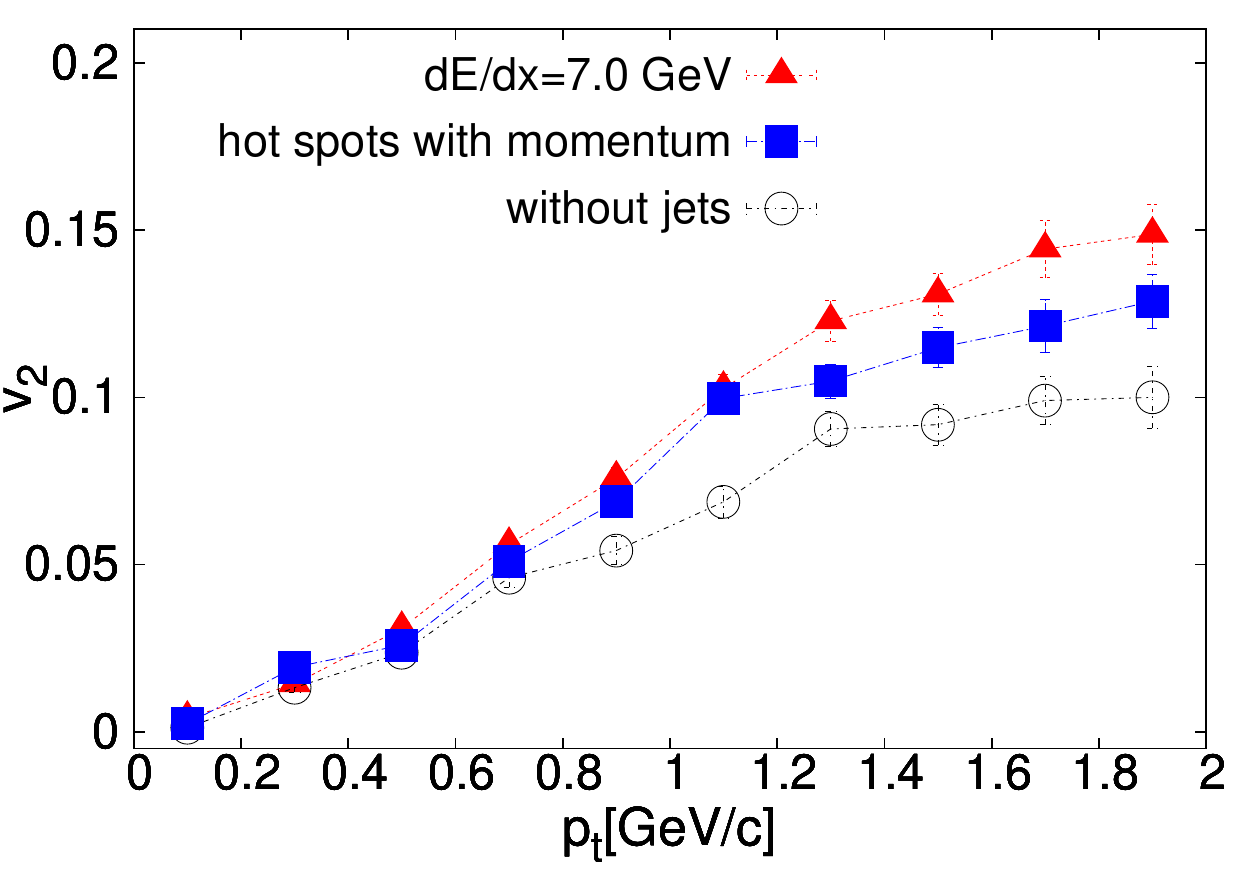}
\caption{Anisotropic flow coefficients generated by momentum deposition from hard partons in centrality class 30--40\%.
Left: $v_2$ and $v_3$ from simulations with momentum deposition  and without it, only with smooth initial conditions
(``without jets''). Right: The curve with squares shows results from a simulation with no momentum deposition from hard partons, 
but with the same amount of momentum and energy anisotropies superimposed on the smooth initial energy density profile in form 
of hot spots. }
\label{f:jets}       
\end{figure*}

Figure~\ref{f:jets}
shows the effect of the momentum depostion. In non-central collisions the elliptic flow is increased by about 50\%
in comparison to simulations with smooth initial conditions. Since there is no triangular anisotropy present in the 
initial conditions, the third-order coefficient $v_3$ is entirely due to energy and momentum deposition from hard 
partons. 

We have also investigated if the effect can be fully included into the initial conditions. To this end, we superimposed 
on the smooth initial energy density profile a number of ``hot spots'' which carried exactly the energy and momentum 
of the hard partons. The resulting $v_2$ (Fig.~\ref{f:jets} right) falls short of the elliptic anisotropy due to hard partons. 

We conclude that momentum deposition from hard partons \emph{during} the evolution of the fireball is an important 
ingredient which must not be omitted in simulations which aim at extracting matter properties via a comparison with 
experimental data \cite{Tachibana:2017syd}.


\section{Event Shape Sorting}
\label{s:ESS}

Anisotropies of the final state distributions of hadrons are to a large extent characteristic for each collision
event. In a single event, it is also possible to measure the anisotropies which disappear in event  averaging, 
e.g.\ by making use of correlation between hadrons. Nevertheless, the question remains if one can select events 
for the analysis in such a way, that at least a part of the characteristics is not washed away in the sum of many 
events. 

Indeed, such a technique has been suggested under the name Event Shape Engineering (ESE) \cite{Schukraft:2012ah}.
When using ESE, first an observable is defined according to the which a selection of events will be done. That observable 
is measured in every event. Events with the value of the observable  within a specified interval are selected into one 
class. 

For example, one determines $|q_2| = |\sum_{i=1}^M e^{2i\phi_i}|/\sqrt{M}$, where $M$ is the multiplicity in the acceptance interval 
and $\phi_i$ is the measured azimuthal angle of the $i$-th recorded hadron.  The events with large $q_2$ are treated 
separately from small $q_2$. ESE is able to separate events effectively and in a well controlled way. 

On the other hand, the choice of the sorting observable (i.e.\ $q_2$ in the above example) must be provided by hand. If there 
exists some hidden structure according to which the events differ in their shapes, it might remain unobserved in 
ESE. In other words, ESE reveals the differences between events, provided that you tell at the beginning how the 
differences look like. 

In order to overcome this feature, we proposed a different method recently \cite{Kopecna:2015fwa,Lehmann:2007pv}, 
under the name Event
Shape Sorting (ESS). In ESS, no observable is pre-defined, according to which the selection of events is made. Instead, the algorithm 
itself re-orders the analysed events in such a way, that events with similar shapes end up close to each other. 

Technically, it works with the histograms in azimuthal angles for individual events. The totality of events is divided into several percentiles, 
typically deciles. We shall refer to these percentiles as to \emph{event bins} and number them from 1 to 10. 
Then, Bayesian probability is determined for each event that it belongs to a given event bin. 
This probability is determined for every event bin. Based on these probabilities the events are then re-arranged in such a 
way, that each event comes to the place where it currently fits best. The procedure is iterated until the order 
of events stops changing in the next iteration. 


\subsection{Results from Event Shape Sorting}
\label{ss:ESSres}

We present here results which we obtained with event sets simulated by two different event generators. 
The first set is composed of $1.5\times 10^5$~events generated by DRAGON \cite{Tomasik:2008fq}. This 
is an MC generator of final state hadrons according to the Blast-Wave model with included resonances. We included 
anisotropies  of second and third order\footnote{%
Anisotropy parameters as formulated in \cite{Cimerman:2017lmm} have been set to 
$a_2,\, \rho_2 \in [-0.1,0.1]$ and $a_3,\, \rho_3\in [-0.03,0.03]$.
} in both shape and flow profile \cite{Cimerman:2017lmm}.

The other two sets of events are provided by AMPT \cite{Lin:2004en}, 
which has been run for Au+Au collisions at impact parameter 7-10~fm at both 
$\sqrt{s_{NN}} = 200$~GeV (AMPT-RHIC) and $\sqrt{s_{NN}} = 2760$~GeV (AMPT-LHC).
There are $10^4$ events in each of the AMPT sets. 

As a first example of how the method works we compare the sorting with the two main 
components of the flow anisotropy in the game: $v_2$ and $v_3$. Note that before 
comparing the shapes of the events we have rotated them in such a way that 
all second-order event planes point in the same direction. In order to be able to follow
a change of the third-order event plane direction, we allow also for negative values of $v_3$. 
They would turn positive if the event plane would be shifted by $\pi/3$.



\begin{figure*}
\centering
\includegraphics[width=0.955\textwidth,clip]{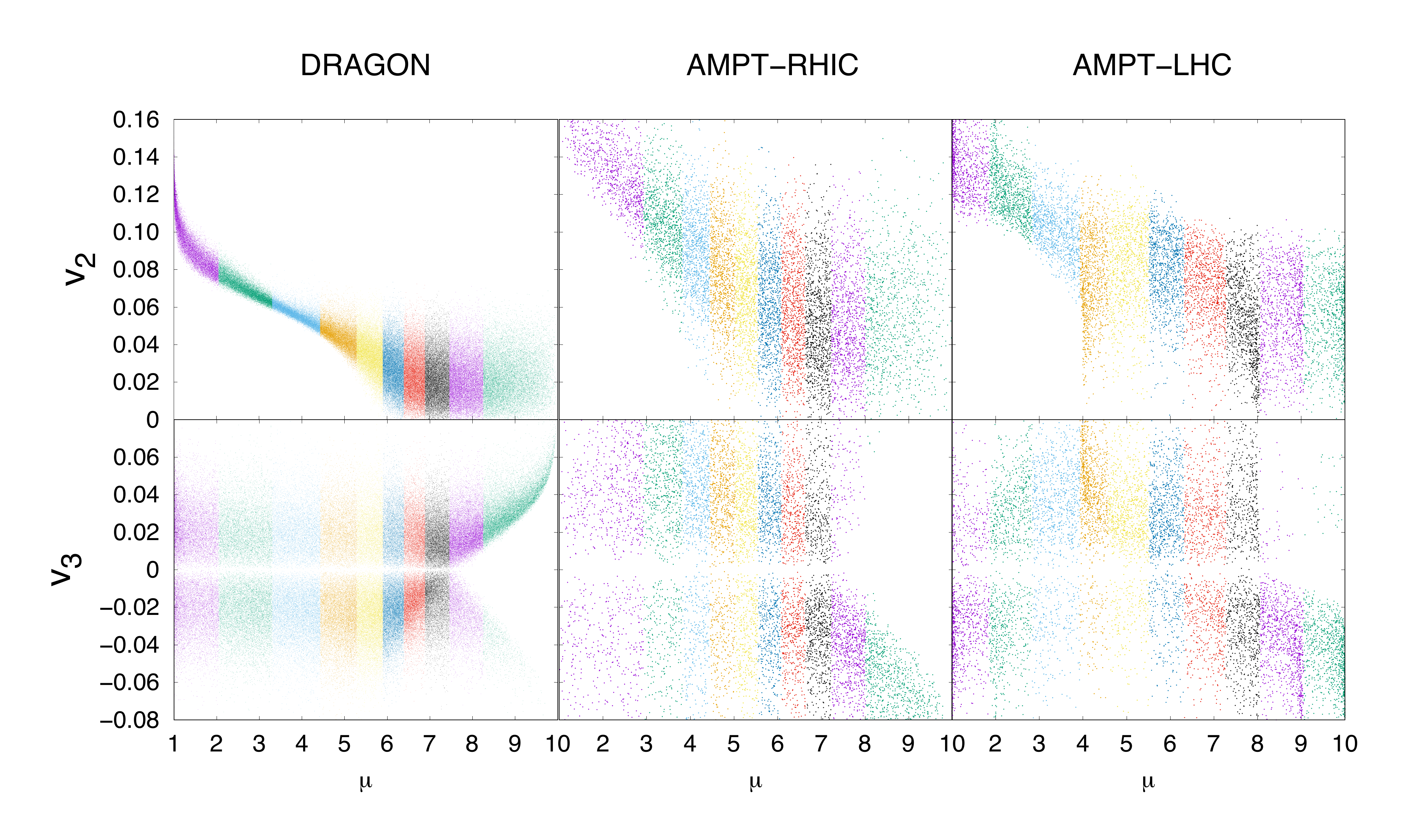}
\caption{Correlation of the sorting variable $\mu$ with $v_2$ (upper row) and $v_3$ (lower row) for all events from the sets generated by DRAGON
(left column), AMPT for RHIC (middle) and AMPT for the LHC (right). The variable $\mu$ 
is the one according to which value the events are ordered.
Different event bins are indicated with different colours; each dot corresponds to one event. }
\label{f:v2v3}       
\end{figure*}

In Figure~\ref{f:v2v3} it is demonstrated how the dominance of the elliptic flow
shows up in the sorting of all events. This is best seen in the set generated by DRAGON. 
The event bins at low sorting variable $\mu$ are determined by large elliptic flow. 
On the other hand, those event bins at high $\mu$'s have all elliptic anisotropy 
within roughly the same interval, but they differ by the triangular anisotropy. 
Qualitatively, similar features are present in the AMPT events, although the magnitude of the 
anisotropies is somewhat higher. Small statistics prevents drawing more precise conclusions from 
the AMPT simulations.

We have also determined the HBT correlation radii, which appear in the Gaussian parametrisation 
of the correlation function 
\begin{equation}
C(q,K) -1 = \exp \left (- R_o^2(K) q_o^2 - R_s^2 (K) q_s^2 - R_l^2 (K) q_l^2 \right )\,  , 
\end{equation}
where $K$ is average pair momentum and the $q$'s are components of the momentum difference (and 
cross-terms vanish in symmetric systems at midrapidity). 
Let us note that they measure the sizes of homogeneity regions, i.e.~those parts of the 
fireball which produce hadrons with specified momentum. 
Due to flow gradients, in an expanding fireball the homogeneity regions are smaller than 
the whole fireball. 
The correlation radii depend on the azimuthal angle, because 
pions flying in different directions come from different homogeneity regions. 

\begin{figure*}
\centering
\includegraphics[width=0.495\textwidth,clip]{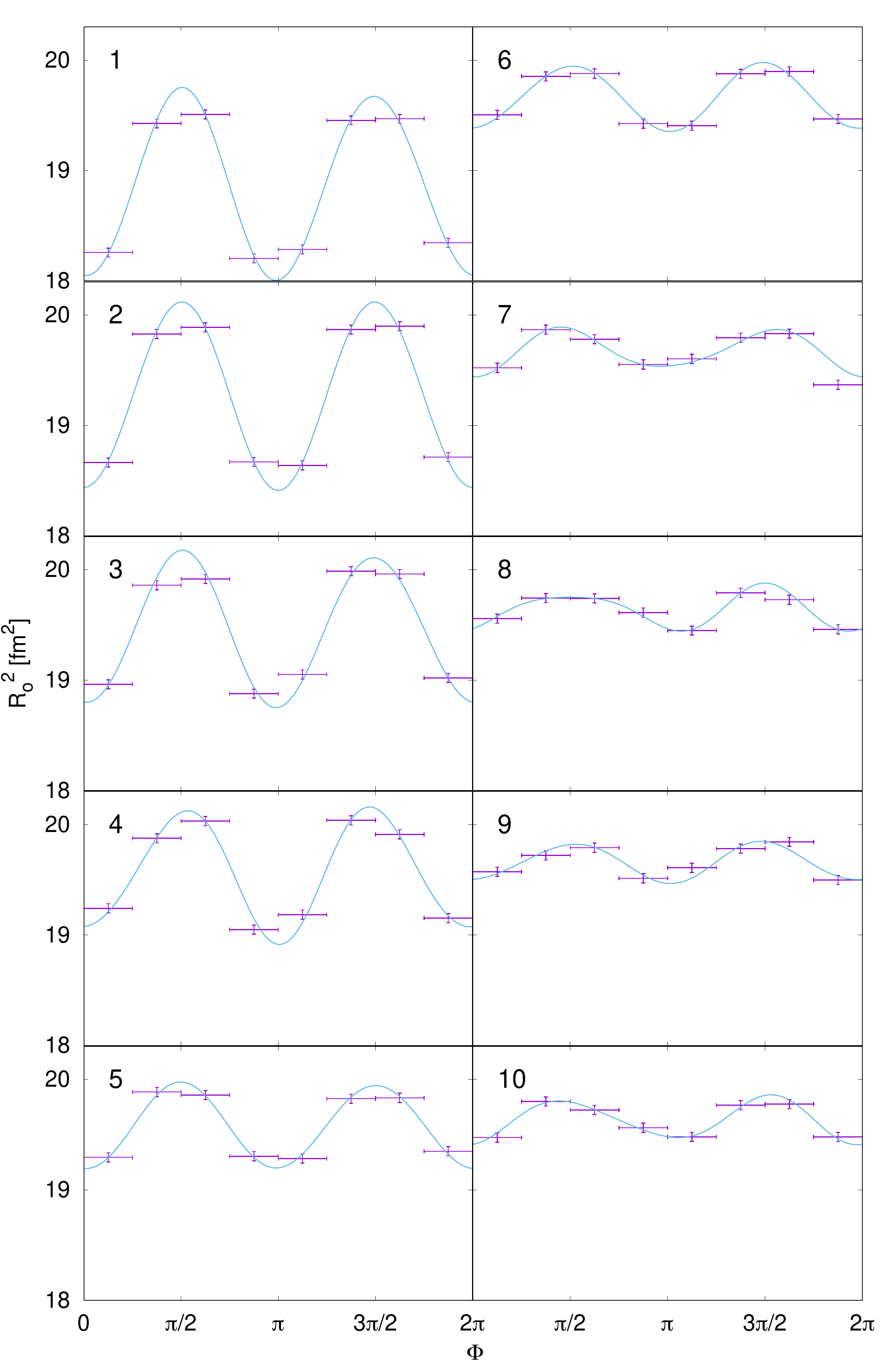}
\includegraphics[width=0.495\textwidth,clip]{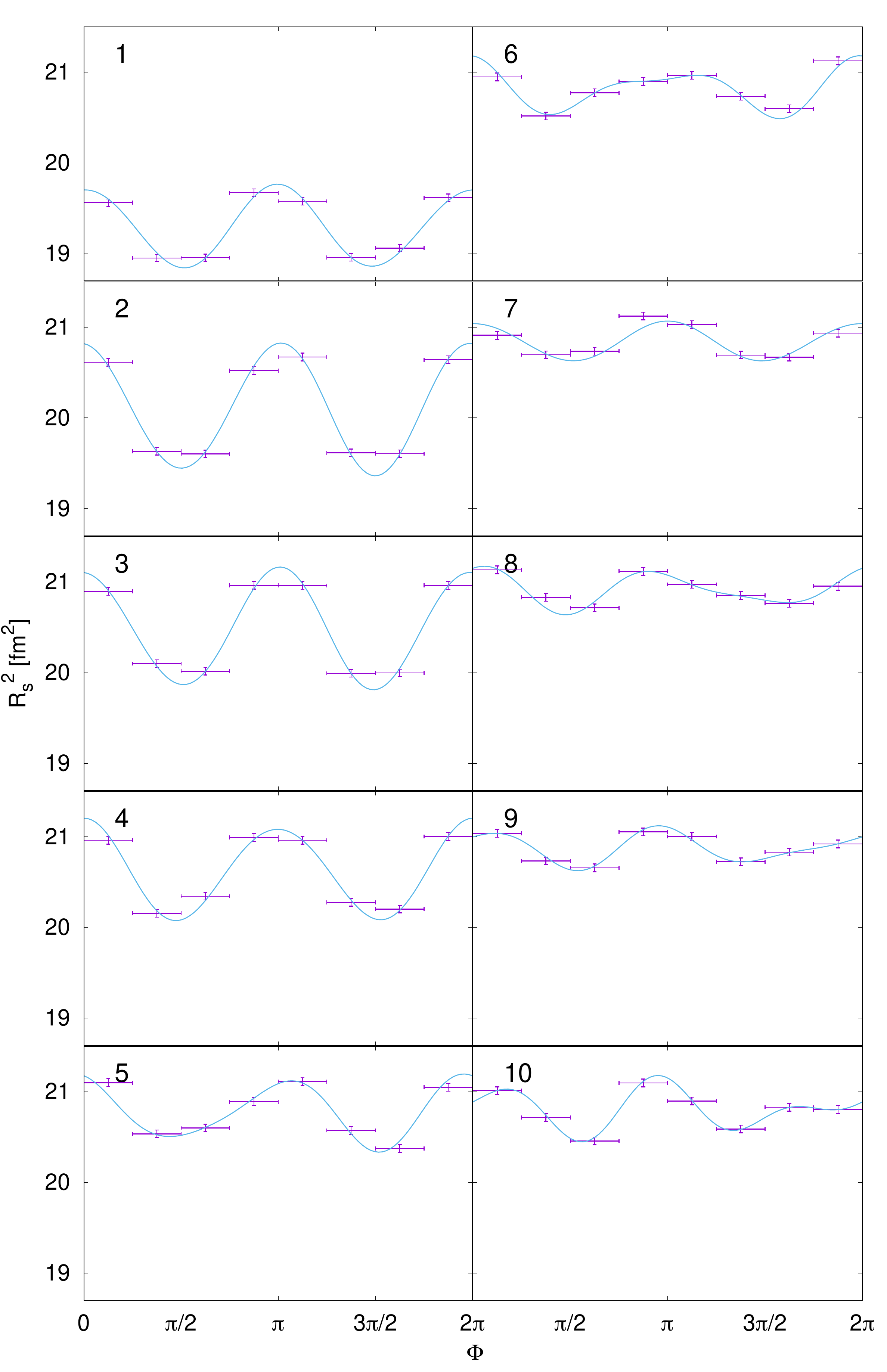}
\caption{Dependences of the correlation radii on the azimuthal angle, as they result from simulations 
with DRAGON. Plotted are $R_o^2$ (left) and $R_s^2$ (right) and different panels show  different event bins, 
as indicated by event bin number in each panel.}
\label{f:radiiphi}       
\end{figure*}

This is shown in Figure~\ref{f:radiiphi}. 
We can see, in accord with Figure~\ref{f:v2v3}, that at lower $\mu's$ there are 
mainly second-order oscillations. The third order Fourier component of the azimuthal 
dependence is seen in event bins with larger $\mu$, also in accord with 
Figure~\ref{f:v2v3}. We stress here, that a measurement of the second and third
order anisotropy in one curve is only possible with the help of ESS. 
They are not seen both together if all events have been just aligned with 
respect to either second or third-order event plane and summed 
up.

\begin{figure*}
\centering
\includegraphics[width=0.955\textwidth,clip]{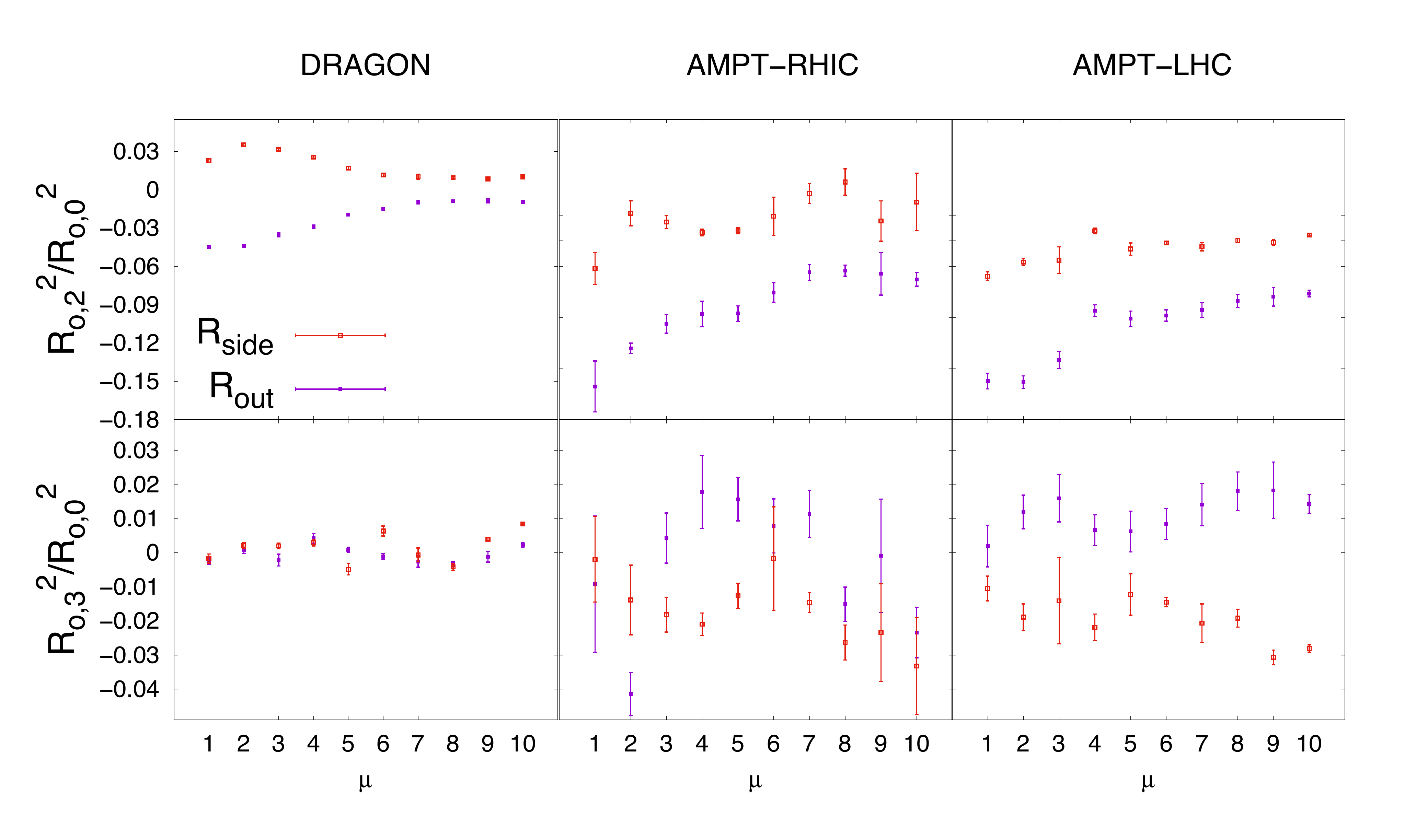}
\caption{Scaled oscillation amplitudes for $R_o^2$ and $R_s^2$ in the second order (upper row)
and the third order (lower row) for different event bins (they are  plotted as functions of 
the event bin number). Simulations by DRAGON (left), AMPT for RHIC (middle) and AMPT for the LHC
(right).}
\label{f:cor_amp}       
\end{figure*}

In order to single out the amplitude of the anisotropies from the overall size of the radii, 
the terms can be scaled by the zeroth-order term
\begin{equation}
R_i^2 = R_{i,0}^2 + \sum_{j=1}^\infty R_{i,j}^2 \cos \left ( n (\phi -\phi_n)\right ) =
R_{i,0}^2 \left ( 1 + \sum_{j=1}^\infty \frac{R_{i,j}^2}{R_{i,0}^2} \cos \left ( n (\phi -\phi_n)\right ) \right )\, .
\end{equation}
The scaled amplitudes are plotted in Figure~\ref{f:cor_amp}. Those from DRAGON simulations actually 
parametrize the curves seen in Figure~\ref{f:radiiphi}. 
The second-order amplitude for DRAGON events 
behaves similarly as $v_2$ in Figure~\ref{f:v2v3}, while the sizes of the third-order amplitude
are small and basically consistent with 0. Larger amplitudes at both orders are seen in the AMPT events.


\section{Conclusions}

Event Shape Sorting provides new insights in the final state distribution of hadrons, since it makes 
it possible to observe both second and third-order oscillations of hadron distributions and correlation radii 
together at the same time. This should allow for a more differential comparison of theory to data. 

We have also argued, that in collisions at the LHC, it is important to add to hydrodynamic 
modelling the momentum deposition from 
hard partons into the bulk medium.


\subsection*{Acknowledgment}
Supported by the grant 17-04505S of the Czech Science Foundation (GA\v{C}R).
BT also acknowledges support by VEGA 1/0348/18 (Slovakia). 

\end{document}